\documentclass[conference]{IEEEtran}
\IEEEoverridecommandlockouts
\usepackage{amsmath,amssymb,amsfonts}
\usepackage{mathptmx}
\usepackage{algorithmic}
\usepackage{graphicx}
\usepackage[numbers,sort&compress]{natbib}
\usepackage{hyperref}
\makeatletter
\let\old@ps@headings\ps@headings
\let\old@ps@IEEEtitlepagestyle\ps@IEEEtitlepagestyle
\def\confheader#1{%
\def\ps@headings{%
\old@ps@headings%
\def\@oddhead{\strut\hfill#1\hfill\strut}%
\def\@evenhead{\strut\hfill#1\hfill\strut}%
}%
\def\ps@IEEEtitlepagestyle{%
\old@ps@IEEEtitlepagestyle%
\def\@oddhead{\strut\hfill#1\hfill\strut}%
\def\@evenhead{\strut\hfill#1\hfill\strut}%
}%
\ps@headings%
}
\makeatother

\confheader{%
4th IEEE International Conference on Recent Advances in Information Technology (RAIT 2018), Mar. 15-17, IIT(ISM) Dhanbad, INDIA
}

\usepackage[pscoord]{eso-pic}
\newcommand{\placetextbox}[3]{
\setbox0=\hbox{#3}
\AddToShipoutPictureFG{ \put(\LenToUnit{#1\paperwidth},\LenToUnit{#2\paperheight}){\vtop{{\null}\makebox[0pt][c]{#3}}}
}
}
\placetextbox{.2}{0.055}{978-1-5386-3038-9/18/\$31.00~\copyright 2018 IEEE}
\begin{document}

\title{Achievable Rate Analysis of Relay Assisted Cooperative NOMA over Rician Fading Channels
} 
\author{\IEEEauthorblockN{Pranav Kumar Jha$^{*}$ and {D. Sriram Kumar$^{}$}}
\IEEEauthorblockA{\textit{$^{}$Department  of Electronics and Communication Engineering},\\ {\textit{National Institute of Technology, Tiruchirappalli, Tamil Nadu, India}} \\
$^{*}$\textit{E-Mail} : jha\_k.pranav@live.com}
}
\maketitle
\begin{abstract}
Non-orthogonal multiple access (NOMA) is a key to multiple access techniques for the next generation 5G wireless communication networks. In this paper, to improve the performance gain of NOMA system, a cooperative fixed decode-and-forward (DF) relay system model based on NOMA (CRS-NOMA) is studied over Rician fading channels, considering the achievable rate of signals as the performance metric. In this technique, by exploiting the concept of NOMA, unlike the conventional method of cooperative relaying, the second time slot is also utilized for the realization of information sent from the transmitter end. Moreover, as the data symbols are transmitted by nodes with full power, the compulsion of complex power allocation coefficients is precluded. In conventional cooperative relaying systems, the receiver is only able to receive a single bit of information, while the receiver can reliably bring in two data symbols in two-time slots. In this regard, this scheme is able to acquire higher achievable rate performance than existing cooperative relaying schemes for larger channel powers over most of the transmit SNR regime. Furthermore, a mathematical expression is also derived for the total achievable rate of CRS-NOMA. The results are verified through Monte-Carlo simulations which validate accuracy and consistency of the derived analytical results.
\end{abstract}
\IEEEpeerreviewmaketitle
\section{Introduction}
\label{sec1}
Non-orthogonal multiple access (NOMA) is the prominent technique and has become a significant basis for the structural composition of radio access schemes for the next generation (5G) wireless communication networks \cite{ding2017application}. However, various multiple access schemes for the 5G have been proposed in literature, including power domain NOMA \cite{saito2013non}, pattern division multiple access (PDMA) \cite{dai2014successive}, sparse code multiple access (SCMA) \cite{taherzadeh2014scma}, and lattice partition multiple access (LPMA) \cite{fang2016lattice} and low density spreading (LDS) \cite{mohammed2012performance}, these schemes depend on the same basic principle, where multiple users are served in each orthogonal channel block. To entertain multiple users, NOMA is completely different from the conventional orthogonal multiple access (OMA) techniques such as time division multiple access (TDMA), frequency division multiple access (FDMA) and orthogonal frequency division multiple access (OFDMA). It can deliver services to multiple users operating with the same channel resources in the time, frequency and spreading code domain, but with the allocation of different power levels, which indicates the multiplexing of multiple user's symbols with superposition coding in the power domain \cite{dai2015non}.
At the receiving end, firstly, decoding of the best quality symbols is achieved and then by applying successive interference cancellation (SIC), detection of the remaining symbols is performed. Moreover, to boost the transmission reliability, network coverage and achievable rate of mobile networks, cooperative relay transmission was proved to be the promising technique \cite{laneman2004cooperative}. In this context, the assimilation of NOMA and cooperative relaying technique has been seen as the assuring approach with high potential and brought intense interests in improving the throughput for the future 5G wireless communication networks in the upcoming years which would certainly be able to fulfill the massive connectivity and demands of wireless networks worldwide. 
\subsection{Related Works}
A cooperative relay assisted NOMA transmission technique was introduced in \cite{ding2015cooperative}, where the users with good quality of channel conditions can be utilized as relays to boost the performance of the users with low quality of channel conditions. In \cite{kim2015non}, NOMA with a cooperative relay was introduced to develop the transmission reliability for a user with imperfect channel conditions. Furthermore, a cooperative relaying system using NOMA (CRS-NOMA) was studied in \cite{kim2015capacity}, where, in the second time slot, only the relay forwards the decoded symbol to the destination. However, most of the existing schemes studied so far, have considered sending the source symbols in the first time slot only, which fails to fully exploit the principle of NOMA for further improvement in the performance of cooperative relaying techniques. In a most recent work in \cite{zhang2017performance}, a different CRS-NOMA scheme is proposed, where for exploiting the NOMA principle at it's fullest and to enhance the performance gain further, second time slot is also used to send the source symbols without the use of complex power allocation coefficient to realize higher achievable rates under Rayleigh fading channels.  Here, in the first time slot, the source transmits the same symbol to both, the relay and the destination. In the second time slot, unlike the conventional-NOMA scheme \cite{kim2015capacity}, the source can transmit symbols to the destination, and the destination can use the SIC detection technique to decode the symbols. In this way, this scheme is achieving higher achievable rates than conventional-NOMA \cite{kim2015capacity} and CRS-OMA \cite{laneman2004cooperative} schemes. Furthermore, as each node transmits the data symbol with its maximum power, the requirement for complicated power allocation is omitted.
\subsection{Motivation and Contribution}
In literature, nearly all the techniques were studied in cooperative NOMA transmission
systems only under Rayleigh fading channels to improve the spectral efficiency \cite{kim2015capacity} and there is no major contribution towards the consideration of Rician fading channels which can proved to be a better candidate with some of the typical 5G wireless network applications. This has strongly influenced the proposal of this work in the current scenario.

In this paper, we analyze the NOMA-based CRS scheme (CRS-NOMA) presented in \cite{zhang2017performance} to fully exploit the NOMA principle and to improve the performance gain over Rician fading channels but at the cost of complex mathematical calculations. We also provide a mathematical expression for the total achievable rate of CRS-NOMA, the consistency of which has been verified using Monte Carlo simulations. Finally, it is shown through the plots generated from simulations that this scheme achieves higher achievable rate than the relaying system based on conventional-NOMA. The simple notation of CRS-NOMA has been used to indicate the cooperative relaying system based on NOMA, which is under investigation.
\subsection{Structure}
The rest of the work is organized in this fashion: in Section \ref{sec2}, a detailed description of the CRS-NOMA system is given along with the equations which provide
its received signals and signal-to-noise ratios (SNRs). Section {\ref{sec3}} presents the mathematical analysis for calculating the total achievable rate for CRS-NOMA,  whereas section \ref{sec4} provides a final expression for it. In Section \ref{sec4}, the derived analytical results are validated numerically and Section \ref{sec5} concludes the work.
\section{System Description}
\label{sec2}
 \begin{figure}
\includegraphics[width=\linewidth]{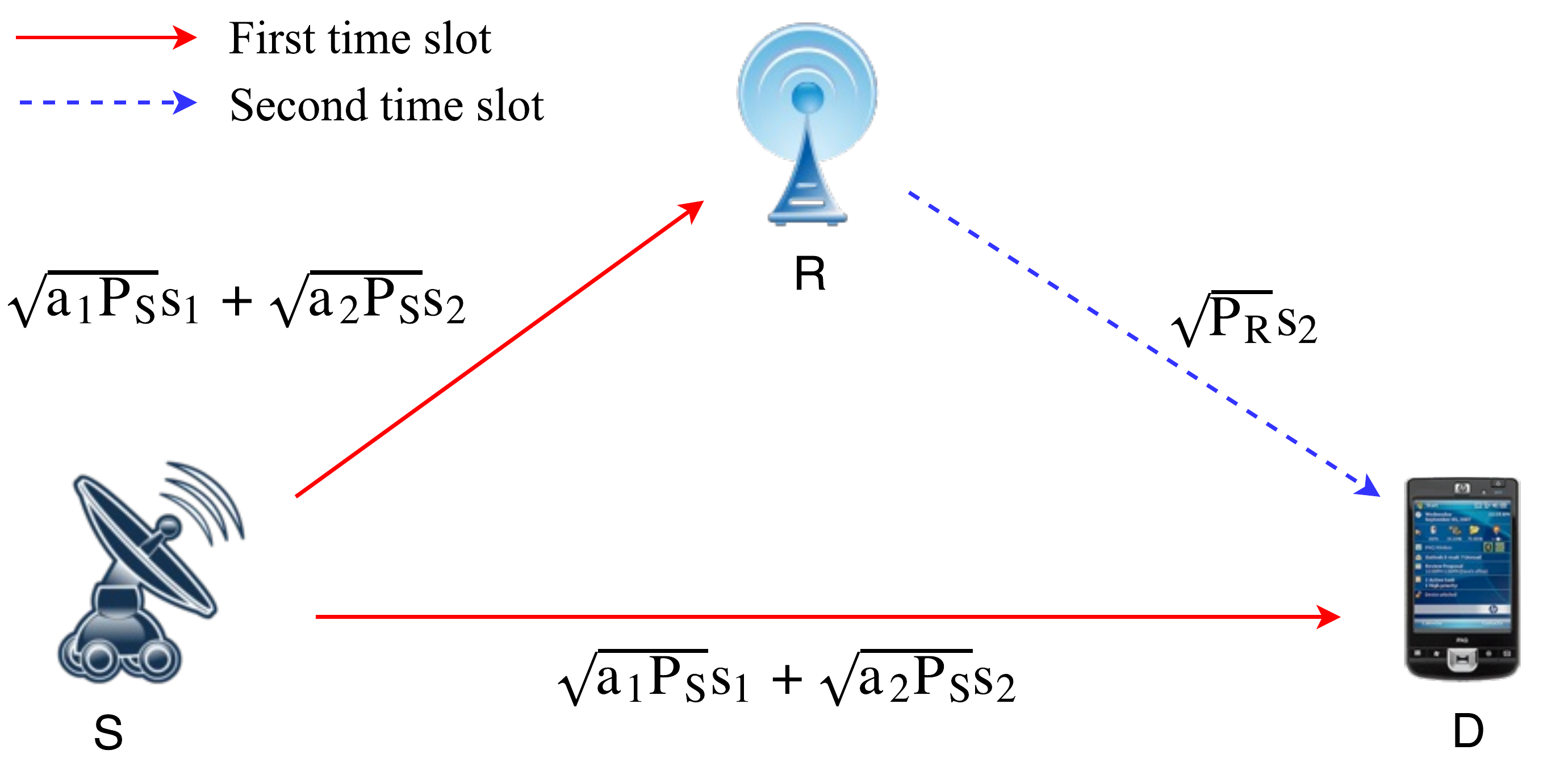}
\caption{A half-duplex DF Relay conventional CRS-NOMA which sends the source symbols in the first time slot and in the second time slot, only the relay forwards the decoded symbols to the destination.}
\label{Fig1}
\end{figure}
\begin{figure}
\includegraphics[width=\linewidth]{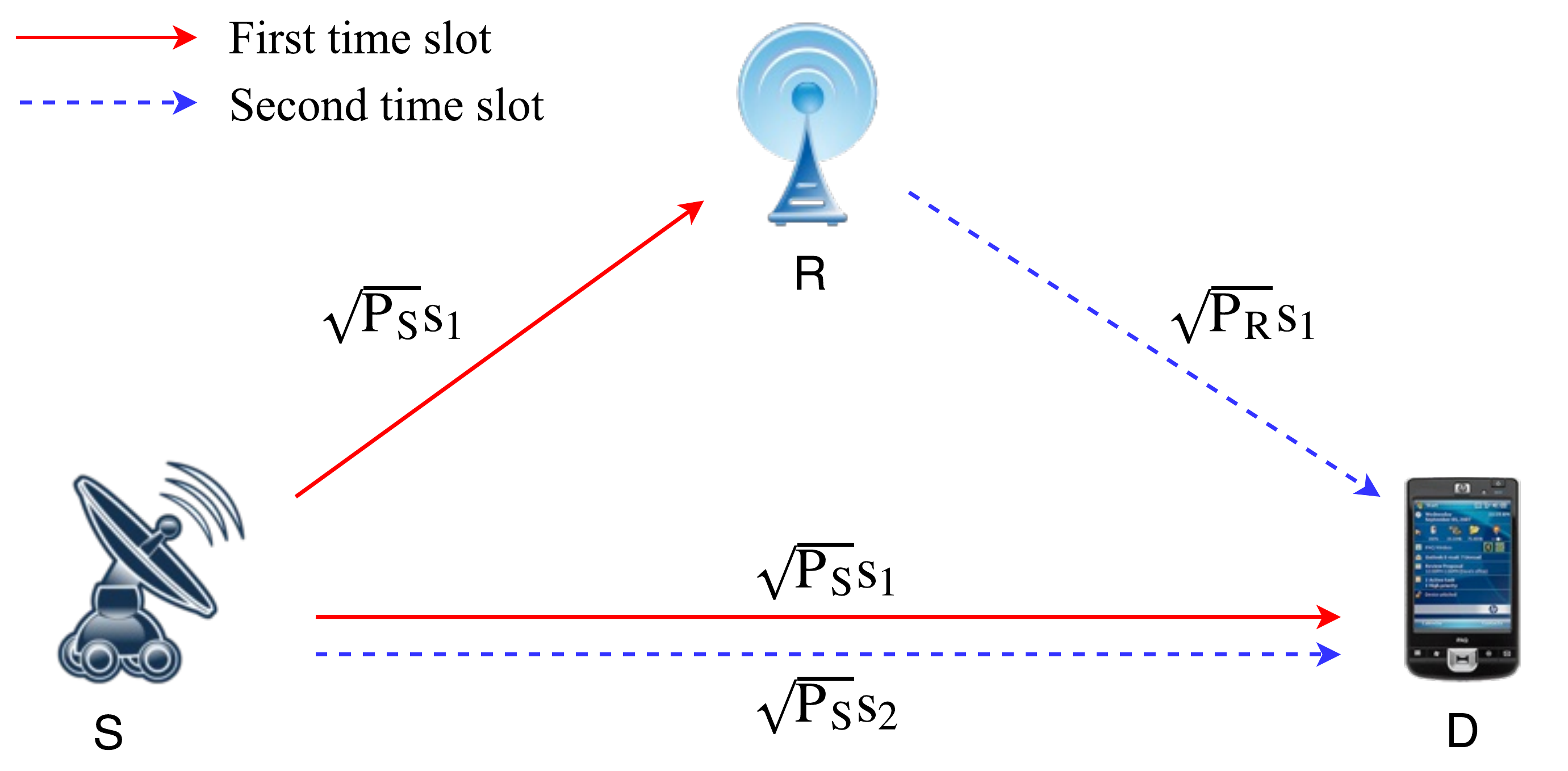}
\caption{A half-duplex fixed DF Relay CRS-NOMA which uses the first as well as the second time slot also to send the source symbols to the destination without complex power allocation coefficients.}
\label{Fig2}
\end{figure}

Fig. \ref{Fig1} illustrates the ordinary cooperative relaying system (CRS), which consists of a source (S), a half-duplex decode-and-forward (DF) relay (R) and a destination (D). 
As depicted in Fig. \ref{Fig2}, a simple cooperative network with source (S), communicates with a destination (D), by means of a fixed DF relay node \cite{laneman2004cooperative} (R), which fully decodes, re-encodes, and retransmits the source message to D. Moreover, S can also directly communicate to D. In addition, it is also assumed that R works in the half-duplex mode which makes it unable to transmit and receive a symbol at the same time. All links between S-R, R-D and S-D are considered to be available and ready to be used. 
The considered independent Rician random variables $h_{SR}$, $h_{RD}$ and $h_{SD}$ are the channel coefficients of S-R, R-D and S-D links with the average powers of $\Omega^2_{SR}$, $\Omega^2_{RD}$ and $\Omega^2_{SD}$, respectively. It is also considered that $\Omega^2_{SR} > \Omega^2_{SD}$, as the path loss of the S-R link, is normally greater than that of the S-D link \cite{zhang2015unified}. The communication process for cooperative relaying systems be composed of two consecutive time slots. In the first time slot, S transmits $s_1$ with power $P_S$ to both R and D. Accordingly, the received signals $r_{SR}$ and $r_{SD}$ at R and D in the first time slot, respectively, are written as,
\begin{equation}
  r_{SR}=\sqrt{P_S}h_{SR}s_1+n_{SR}
  \label{eq1}
\end{equation}
\begin{equation}
  r_{SD}=\sqrt{P_S}h_{SD}s_1+n_{SD}
   \label{eq2}
\end{equation}
where $n_{SR}, n_{SD}\sim\mathcal{N}(0,\sigma^2)$ which represents zero mean additive white Gaussian noise (AWGN) with variance $\sigma^2$. Here, $s_i$ represents the $i$-th data symbol, $P_S$ signifies the total transmit power from S and $E[{|s_i|}^2]=1$ with $E\{\cdot\}$ denoting expectation. The received SNR for $s_1$ at R can be written as,
\begin{equation}
  \gamma_{SR},s_1=\frac{P_S|h_{SR}|^2}{\sigma^2}= \rho |h_{SR}|^2
   \label{eq3}
\end{equation}
and the received SNR for $s_1$ at D is
obtained as,
\begin{equation}
  \gamma_{SD},s_1=\frac{P_S|h_{SD}|^2}{\sigma^2}= \rho |h_{SD}|^2
   \label{eq4}
\end{equation}
In the second time slot, R forwards decoded signal $s_1$ with power $P_R$ to D and S transmits $s_2$ with power $P_S$ to D. While, in the CRS-NOMA scheme \cite{kim2015capacity}, only R transmits the decoded signal to D as in Fig. \ref{Fig1}. Then, the received signal at D is given by,
\begin{equation}
  r_{RD}=\sqrt{P_R}h_{RD}s_1+\sqrt{P_S}h_{SD}s_2+n_{RD}
   \label{eq5}
\end{equation}
where $n_{RD}\sim\mathcal{N}(0,\sigma^2)$ and $P_R$ denotes the total transmit power from R. Generally, the fading gain $h_{SD}$ of the S-D link is smaller than the fading gain $h_{RD}$ of the R-D link represented by $E\{|h_{SD}|^2\} < E\{|h_{RD}|^2\}$, as a result of differences in the distances of S and R from D, respectively. This characteristic of the channel difference between different transceiver pairs makes it easier to utilize the NOMA principle at the second time slot. However, CRS-NOMA in \cite{kim2015capacity} employs complex power allocation coefficients $a_1$ and $a_2$ at S to distinguish two signals. Analogous to the SIC-based NOMA technique, D firstly decodes $s_1$ by treating $s_2$ as a noise term. Then, $s_1$ is canceled from $r_{RD}$ by using SIC to decode $s_2$. It is considered that $P_S = P_R = P$, where $P$ represents the total transmit power. Therefore, the received SNRs of $s_1$ and $s_2$ can be written as,
\begin{equation}
  \gamma_{RD},s_1=\frac{\frac{P|h_{RD}|^2}{\sigma^2}}{\frac{P|h_{SD}|^2}{\sigma^2}+1}
  =\frac{P|h_{RD}|^2}{P|h_{SD}|^2+\sigma^2}
   \label{eq6}
\end{equation}
\begin{equation}
  \gamma_{SD},s_2 = \frac{P|h_{SD}|^2}{\sigma^2} = \rho |h_{SD}|^2 
   \label{eq7}
\end{equation}
\section{Achievable Rate Analysis}
\label{sec3}
According to the concept of NOMA and using the fixed DF relay technique proposed in \cite{zhang2017performance}, the total achievable rate of $s_1$ is given as,
\begin{equation}
\begin{split}
C_{S1} & = \underbrace{\frac{1}{2}\textrm{ min }\{\textrm{log}_2(1+\gamma_{RD},s_1),\textrm{log}_2(1+\gamma_{SR},s_1)}_{C_{R,{s_1}}}\}\\
 & + \underbrace{\frac{1}{2}\textrm{log}_2(1+\gamma_{SD},s_1)}_{C_{D,{s_1}}}
\label{eq8}
\end{split}
\end{equation}
where $C_{R,{s_1}}$ denotes the achievable rate of $s_1$ transmitted through R and $C_{D,{s_1}}$ represents the achievable rate of $s_1$ transmitted directly from S-to-D. The parameter 1/2 is present as the two-time slots are used in the relaying system. The achievable rate of $s_2$ is $C_{S2} = C_{D,{s_1}}$, which can be written as, 
\begin{equation}
C_{S2} = C_{D,{s_1}}  = \frac{1}{2}\textrm{log}_2(1+\gamma_{SD},s_1)
\label{eq9}
\end{equation}
Furher, in consequence of (\ref{eq8}) and (\ref{eq9}), the total achievable rate can be calculated as,
\begin{equation}
C=C_{S1}+C_{S2}={C_{R,{s_1}}}+2{C_{D,{s_1}}}\label{eq1212}
\end{equation}
\begin{equation}
\begin{split}
C & = \frac{1}{2}\textrm{ min }\{\textrm{log}_2(1+\gamma_{RD},s_1),\textrm{log}_2(1+\gamma_{SR},s_1)\}\\
 & + \textrm{log}_2(1+\gamma_{SD},s_1) 
\label{eq19}
\end{split}
\end{equation}
Now, 
let $\lambda_{SD}\triangleq|h_{SD}|^2$, $\lambda_{SR}\triangleq|h_{SR}|^2$, $\lambda_{RD}\triangleq|h_{RD}|^2$, $\rho=\frac{P}{\sigma^2}$ and $C(x)\triangleq \textrm{log}_2(1+x)$, where $\rho$
is the transmit SNR. Hence, using (\ref{eq19}), the total achievable rate $C$ of signals $s_1$ and $s_2$  in terms of $\lambda_{x}$, where $x\in\{SR, RD, SD\}$, can be obtained as \cite{kim2015capacity},
\begin{equation}
\begin{split}
C & = \frac{1}{2}\textrm{ min }\{\textrm{log}_2(1+\gamma_{RD},s_1),\textrm{log}_2(1+\gamma_{SR},s_1)\}\\
& + \textrm{log}_2(1+\gamma_{SD},s_1)\\
& = \frac{1}{2}\textrm{log}_2(1+\textrm{min}\{\lambda_{RD}, \lambda_{SR}\}\rho) + \textrm{log}_2(1+\lambda_{SD}\rho)
\label{eq10}
\end{split}
\end{equation}
This expression in (\ref{eq10}) provides the total achievable rate for the system model presented and the final calculated value of which will be used to generate the MATLAB plots against the transmit SNR $\rho$ for different channel powers in Section \ref{sec4}.

\noindent Now, let, $\gamma_1\triangleq \textrm{min}{\{\lambda_{RD},\lambda_{SR}\}}$,  $\gamma_2\triangleq \lambda_{SD}$, the cumulative distribution functions (CDFs) of $\gamma_1$ for $C_{R,{s_1}}$ and $\gamma_2$ for $C_{D,{s_1}}$ can be written as \cite{7983401},
\begin{equation}
\begin{split}
F(\gamma_1) & = 1-A_x A_y\sum_{k=0}^{\infty}\sum_{n=0}^{\infty}\tilde B_x(n)\tilde B_y(k) n!k!e^{-(a_x+a_y)\gamma_1} \\
& \times \sum_{i=0}^{n}\sum_{j=0}^{k} \frac{a_x^ia_y^j}{i!j!}\gamma_1^{i+j}
\label{eq12}
\end{split}
\end{equation}
\begin{equation}
\begin{split}
F(\gamma_2) & = 1-A_z A_y\sum_{k=0}^{\infty}\sum_{n=0}^{\infty}\tilde B_z(n)\tilde B_y(k) n!k!e^{-(a_z+a_y)\gamma_2} 
\\
& \times\sum_{i=0}^{n}\sum_{j=0}^{k} \frac{a_z^ia_y^j}{i!j!}\gamma_2^{i+j}
\label{eq13}
\end{split}
\end{equation}
where, $B_x(n)=K_x^n(1+K_x)^n/\Omega_x^n(n!)^2, B_y(k)=K_y^k(1+K_y)^k/\Omega_y^k(k!)^2, a_x=(1+K_x)/\Omega_x,  a_y=(1+K_y)/\Omega_y, A_x=a_xe^{-K_x}, A_y=a_ye^{-K_y},$
$\tilde B_x(n)=B_x(n)/a_x^{n+1},
\tilde B_y(k)=B_y(k)/a_y^{k+1}$. Here, S-D, S-R and R-D links are represented by $x$, $y$ and $z$, respectively, whereas $K$ represents the Rician factor.
Now, as the use of complex power allocation coefficients is excluded for CRS-NOMA, then $A_x \approx e^{-K_x}, A_y \approx e^{-K_y}$ and (\ref{eq12}) and (\ref{eq13}) can be modified for its corresponding values as,
\begin{equation}
\begin{split}
F(\gamma_1) & \approx 1-A_x A_y\sum_{k=0}^{\infty}\sum_{n=0}^{\infty} B_x(n) B_y(k) n!k!e^{-\gamma_1} \sum_{i=0}^{n}\sum_{j=0}^{k} \frac{\gamma_1^{i+j}}{i!j!}
\label{eqN1}
\end{split}
\end{equation}
\begin{equation}
\begin{split}
F(\gamma_2) & \approx 1-A_z A_y\sum_{k=0}^{\infty}\sum_{n=0}^{\infty} B_z(n) B_y(k) n!k!e^{-\gamma_2} 
\sum_{i=0}^{n}\sum_{j=0}^{k} \frac{\gamma_2^{i+j}}{i!j!}
\label{eqN2}
\end{split}
\end{equation}
According to \cite{7983401}, \eqref{eqN1} and \eqref{eqN2} can be approximated by the asymptotic results where approximated values for $C_{R,{s_1}}$ and $C_{D,{s_1}}$ is given as,
\begin{equation}
C_{R,{s_1}}=\frac{1}{2\textrm{ln}(2)}[H(\rho)]\label{eq14}
\end{equation}
\begin{equation}C_{D,{s_1}}=\frac{1}{2\textrm{ln}(2)}[G(\rho)]
\label{eq31}
\end{equation}
and by using (\ref{eq1212}), (\ref{eq14}) and (\ref{eq31}), the total achievable rate $C$ can be calculated as,
\begin{equation}
C=\frac{1}{2\textrm{ln}(2)}[H(\rho)+2G(\rho)]\label{eq20}
\end{equation}
where $H(\rho)$ and $G(\rho)$ mentioned are given below as,
\begin{equation}
\begin{split}
H(\rho) & \approx A_x A_y\sum_{k=0}^{\infty}\sum_{n=0}^{\infty} B_x(n) B_y(k) n!k!
\sum_{i=0}^{n}\sum_{j=0}^{k} \frac{(i+j)!}{i!j!\rho^{i+j}}\\
& \times e^\frac{1}{\rho}\bigg(2\frac{1}{\rho}\bigg)^{-(i+j)}\frac{\pi}{n}\sum_{t=1}^{n}(\textrm{cos}(\frac{2t-1}{2n}\pi)+1)^{i+j-1}
\\
& \times e^{-\frac{2{\rho^{-1}}}{\textrm{cos}(\frac{2t-1}{2n}\pi)+1}}|\textrm{sin}(\frac{2t-1}{2n}\pi)|
\end{split}\label{eq15}
\end{equation}
\begin{equation}
\begin{split}
G(\rho) & \approx A_z A_y\sum_{k=0}^{\infty}\sum_{n=0}^{\infty} B_z(n) B_y(k) n!k!\sum_{i=0}^{n}\sum_{j=0}^{k} \frac{(i+j)!}{i!j!\rho^{i+j}}
\\
& \times e^\frac{1}{\rho}\bigg(2\frac{1}{\rho}\bigg)^{-(i+j)}\frac{\pi}{n}\sum_{t=1}^{n}(cos(\frac{2t-1}{2n}\pi)+1)^{i+j-1}
\\ & \times e^{-\frac{2{\rho^{-1}}}{\textrm{cos}(\frac{2t-1}{2n}\pi)+1}}|\textrm{sin}(\frac{2t-1}{2n}\pi)|
\label{eq16}
\end{split}
\end{equation}

\section{Numerical Results and Discussions}
\label{sec4}
\begin{figure}
\includegraphics[width=\linewidth]{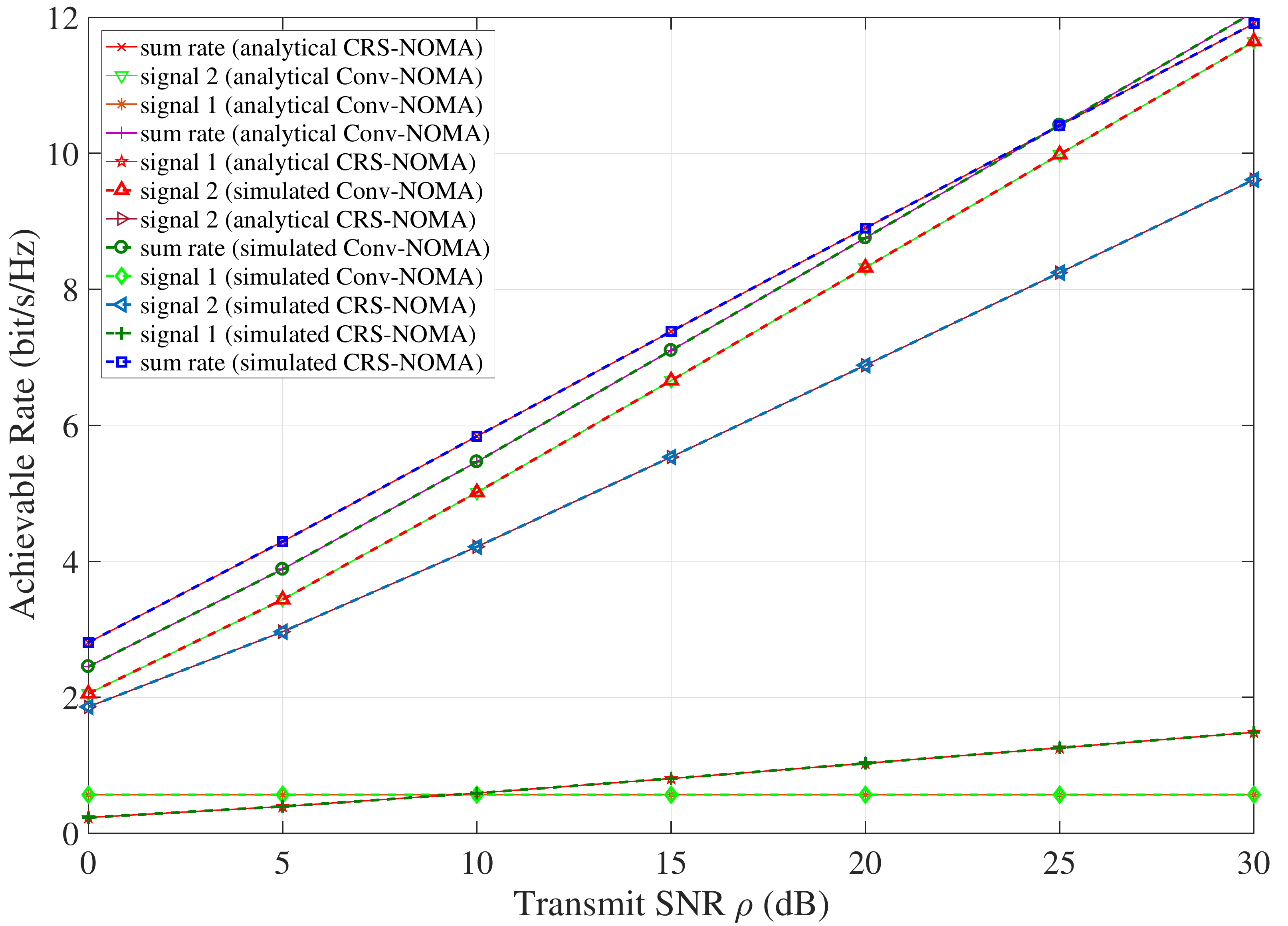}
\caption{Achievable rates of CRS-NOMA and conventional-NOMA relaying systems for $a_1=0.9$ and $a_2=0.1$ when $\Omega_{SD}=3$, $\Omega_{SR}=\Omega_{RD}=8$.}
\label{Fig3}
\end{figure}

\begin{figure}
\includegraphics[width=\linewidth]{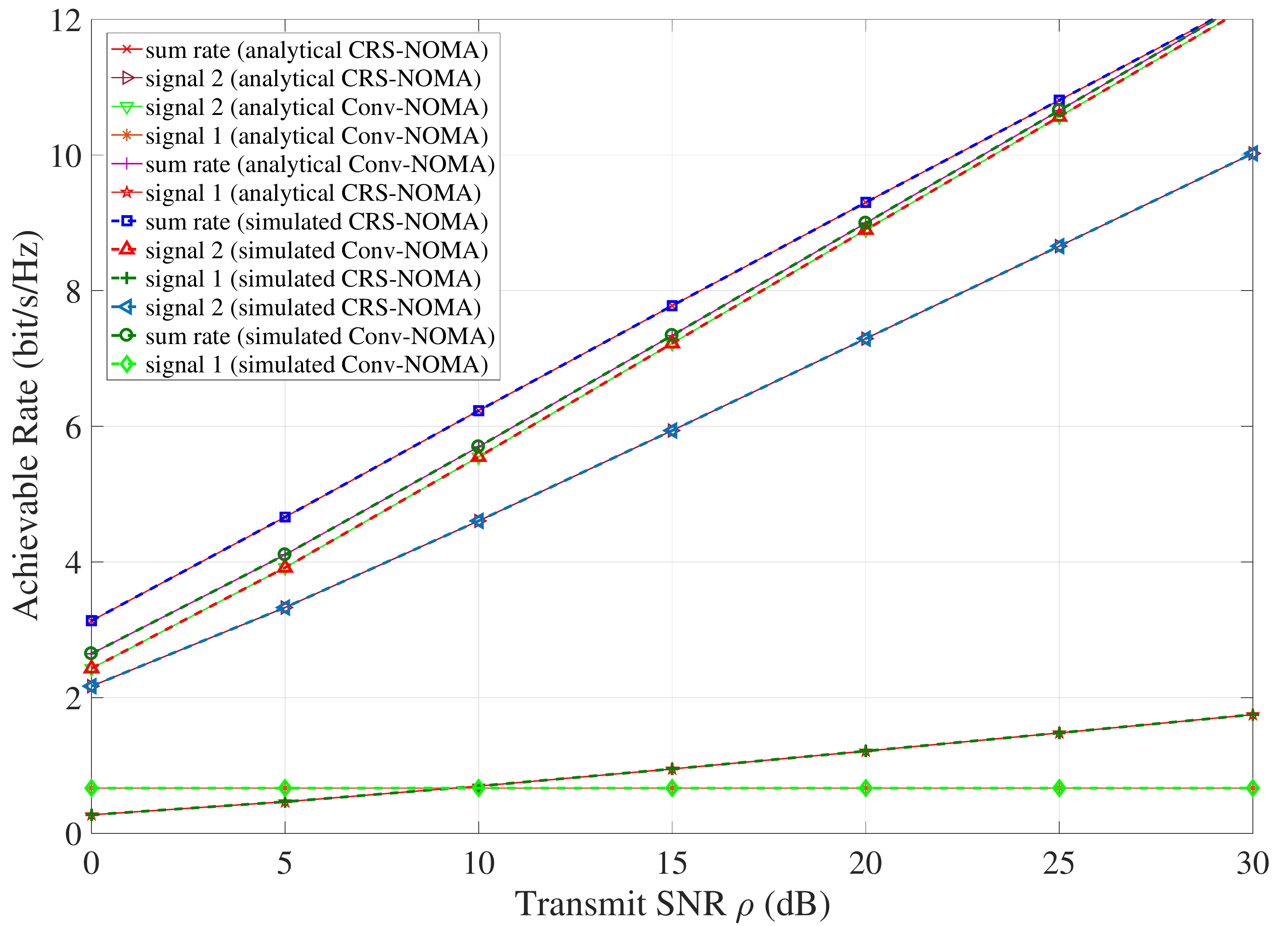}
\caption{Achievable rates of CRS-NOMA and conventional-NOMA relaying systems for $a_1=0.9$ and $a_2=0.1$ when $\Omega_{SD}=3$, $\Omega_{SR}=\Omega_{RD}=12$.}
\label{Fig4}
\end{figure}

Numerical results show that the derived analytical results for CRS-NOMA system model for Rician fading channels in \eqref{eq20} is well matched with the results obtained from Monte-Carlo simulations, which validates the consistency of the analytical curves with the simulation ones. The numerical results are also compared with the conventional cooperative relaying NOMA scheme \cite{7983401} in Figs. \ref{Fig3} and \ref{Fig4} and it is clear from the plots that the derived result for CRS-NOMA outperforms the conventional-NOMA, which became possible owing to the use of SIC at the receiver and enabling the transmission of $s_2$ during the second time slot. With the help of relay node, the fading gain of $s_1$ is also improved by making the R-D distance shorter represented by the larger values of $\Omega_{RD}$ as shown in Fig. \ref{Fig4}.

In Fig. \ref{Fig3}, the channel powers are considered as, $\Omega_{SD}=3$, $\Omega_{SR} = \Omega_{RD} = 8$ with power allocation coefficients as $a_1=0.9$ and $a_2=0.1$. When $\rho$ increases from $5$ dB to $25$ dB, the achievable sum rate of CRS-NOMA increases from $4.29$ bit/s/Hz to $10.41$ bit/s/Hz and from $3.883$ bit/s/Hz to $10.41$ bit/s/Hz for conventional-NOMA, indicating that CRS-NOMA has $0.407$ bit/s/Hz higher achievable rate than conventional scheme of NOMA at $\rho=5$ dB. Further, $s_2$ increases from $2.964$ bit/s/Hz to $8.245$ bit/s/Hz for CRS-NOMA and $3.438$ bit/s/Hz to $9.984$ bit/s/Hz for conventional relaying scheme. At $\rho=25$ dB, $s_1$ is $1.256$ bit/s/Hz for CRS-NOMA and $0.5663$ bit/s/Hz for conventional-NOMA. 

In Fig. \ref{Fig4}, to gain further improvement in the performance of CRS-NOMA, larger channel powers are used as, $\Omega_{SR} = \Omega_{RD} = 12$, whereas $\Omega_{SD} = 3$ which is fixed for the analysis. Now, when $\rho$ increases from $5$ dB to $25$ dB, the achievable sum rate of CRS-NOMA increases from $4.662$ bit/s/Hz to $10.81$ bit/s/Hz and from $4.107$ bit/s/Hz to $10.66$ bit/s/Hz for conventional one, whereas $s_2$ increases from $3.328$ bit/s/Hz to $8.657$ bit/s/Hz for CRS-NOMA and $3.916$ bit/s/Hz to $10.57$ bit/s/Hz for other system. At $\rho=25$ dB, $s_1$ is $1.48$ bit/s/Hz for CRS-NOMA and $0.667$ bit/s/Hz for conventional cooperative relaying NOMA technique. Here, a performance gain of $0.56$ bit/s/Hz for CRS-NOMA over conventional-NOMA is observed at $\rho=5$ dB which is $0.153$ bit/s/Hz more than the rate gain in Fig. \ref{Fig3}. It can also be seen that, at $\rho=25$ dB, CRS-NOMA achieves $10.41$ bit/s/Hz in Fig. \ref{Fig3} and $10.81$ bit/s/Hz in Fig. \ref{Fig4}.

It is clearly demonstrated from the results that CRS-NOMA outperforms the conventional-NOMA in terms of total achievable rate for a transmit SNR value up to 25 dB, but after this, conventional-NOMA starts to dominate this system, which indicates that CRS-NOMA is performing well within a threshold or limit, which can be set at $25$ dB in this case. Next, the signal rate for $s_2$ is superior in the case of conventional-NOMA in contrast to the CRS-NOMA for all values of $\rho$. For signal $s_1$, this is quite complex in nature and shows that signal rate for conventional-NOMA is almost same for the $\rho$ value used but increasing linearly in the case of CRS-NOMA which crosses the signal rate at $\rho=10$ dB and the possible explanation for this behavior might be the omission of complex power allocation for signal transmission of each node transmitting the data symbols with its maximum power. Moreover, use of larger channel powers demonstrate that the rates achieved by CRS-NOMA are always leveling up for most of the transmit SNR regime and also helps in enhancing the spectral efficiency of the signals $s_1$, $s_2$ and in succession, improves the performance gain of CRS-NOMA systems making it the most promising technique for the future.  
\section{Conclusions}
\label{sec5}
In this paper, a fixed DF relay assisted cooperative NOMA scheme is analyzed without the use of complex power allocation by sending the data symbols from source-to-relay and source-to-destination in two different time slots under Rician fading channels. The system model is analyzed and each key point is explained mathematically which provides a clear description of the CRS-NOMA system model. Moreover, an asymptotic mathematical expression is also provided for the achievable rate of CRS-NOMA system and plotted against the transmit SNR $\rho$ where the analytical curves are consistent with the simulation results as shown by the MATLAB figures. The analysis presented in this work shows that CRS-NOMA outperforms conventional-NOMA in terms of achievable rate, which became possible owing to the use of SIC at the receiver and enabling the transmission of data symbols during the second time slot. It is also demonstrated that larger average channel power of the R-D link has a favorable effect on the performance of the system. In the end, it can be concluded that relaying systems based on CRS-NOMA scheme presented has better spectral efficiency than conventional-NOMA relaying systems. Furthermore, it can be stated that NOMA can be integrated with the future 5G wireless systems and other existing systems because of
its compatibility with other communication technologies under various circumstances.

\bibliographystyle{IEEEtran}
\bibliography{references}

\end{document}